\documentclass[preprint,12pt]{aastex}
\usepackage{epsfig}


\def\wdf{white dwarf}
\def\ce{common envelope}
\def\rl{Roche lobe}
\def\m-s{main-sequence}
\def\ms{main sequence}

\def\m-t{mass-transfer}
\def\kep{{\it Kepler}}

\def\tr{transit}
 
\def\det{detect} 

\begin{document} 

\title{
Transits and Lensing by Compact Objects in the {\it Kepler} Field:
Disrupted Stars Orbiting Blue Stragglers}

\author{R. Di\thinspace Stefano
}
\affil{
Harvard-Smithsonian Center for Astrophysics,
60 Garden St..
Cambridge, MA 02138}

\begin{abstract}
{\it Kepler's} first major discoveries are two hot
($T > 10,000$~K) small-radius objects orbiting stars in 
its field.
A viable hypothesis is that these are the cores of stars that have each been
eroded or disrupted by a companion star. 
The companion, which is the star monitored today, is likely to have gained    
mass from its now-defunct partner, and can be 
considered to be a blue straggler.
KOI-81 is almost certainly
the product of stable mass transfer; KOI-74 may be as well, or it may be the
first clear example of a blue straggler created through three-body interactions. 
We show that mass 
transfer binaries are
common enough that 
{\it Kepler} should discover $\sim 1000$ white dwarfs 
orbiting main sequence stars. Most, like KOI-74 and KOI-81,
 will be discovered through transits, but many 
will be discovered through a combination of gravitational lensing and transits,
while lensing will dominate for a subset. In fact,
some events caused by white dwarfs will have the appearance of ``anti-transits''--i.e., short-lived
enhancements in the amount of light received from the
monitored star. Lensing and other mass measurements methods provide
a
way to distinguish white 
dwarf binaries from
planetary systems. This is important for the success of
{\it Kepler's} primary mission, in light of the fact that  
white dwarf radii are similar to the 
radii of terrestrial planets,
and that some white dwarfs will have  orbital periods
that place them in the habitable zones of their stellar companions.
By identifying transiting and/or lensing white dwarfs, 
{\it Kepler} will conduct pioneering
studies of white dwarfs and of the end states of mass transfer. It may
also identify orbiting neutron stars or black holes.
The calculations inspired by the discovery of KOI-74 and KOI-81
have implications for ground-based wide-field surveys as well as for 
future space-based surveys.   
\end{abstract}

\section{Introduction}

The {\it Kepler} mission was 
launched on 6 March 2009.  Its 
most startling and potentially important
results so far are the discoveries of
small, luminous, hot objects orbiting two 
of the stars monitored
by {\it Kepler} (Rowe et al.\, 2010, hereafter R2010).
Each of two ``Kepler Objects of Interest'' 
(KOIs) has exhibited
repeating transits, and intervening eclipses 
that are deeper than the transits. In KOI-81, 
it is estimated
that the luminosity and temperature of the 
compact companion,
KOI-81b,  
are $0.4\, L_\odot$ and $\sim 13,000$K, respectively.
For KOI-74b the estimates are  $L=0.03\, L_\odot$ and 
$T \sim 12,000$K.  
These small objects have temperatures ten times higher 
than would be consistent with heating by the star alone.
We explore the possibility that the hot compact
objects
are the cores of stars whose evolutions were interrupted 
by binary interactions, focusing on the implications for the 
{\it Kepler} mission and for fundamental science. 

In \S 2 we consider the general
 evolutionary sequences that may
have produced KOI-81 and KOI-74. 
We turn in \S 3 to the question of the likelihood of
detecting the products of mass transfer in the {\it Kepler}
data set. We show 
that many more systems that have experienced mass transfer
 are likely to be discovered by {\it Kepler}, making 
the mission a unique tool for the study of
interacting binaries. 
Because the eclipse depths and orbital separations
of some white-dwarf/blue-straggler binaries are similar to what is expected
of terrestrial planets orbiting in the habitable zones of their host stars,
it is important to distinguish between the two possibilities.
Mass measurements are crucial. An interesting consequence of the
  higher mass of white dwarfs relative to planets is that white dwarfs can produce
detectable lensing of the stars they orbit.  The magnification associated with lensing
can balance the dimming associated with a normal transit, or it can
produce a net magnification (Sahu \& Gilliland 2003;
Farmer \& Agol 2003; Agol 2003). We incorporate the
effects of lensing in \S 4 and summarize our results in \S 5. 
 

\section{Erosion and Disruption}
\subsection{Evolutionary sequences}
When a \wdf\ orbits a main-sequence  star, the progenitor of the \wdf\ must have
initially been the most massive star in the binary. Let $M_1(0)$ denote
the initial mass of the primary.
If the mass of the \wdf, $M_c,$ is smaller than that of the remnant normally produced
by a star of this mass, we can infer that the  
evolution of its progenitor was truncated through interaction
with the secondary.
The interaction may have been stable, producing a gradual erosion
of the primary through mass transfer. Or it may have been unstable,
producing an almost explosive disruption of the primary.
In either case,
the mass $M$ of the main-sequence star we observe
today may be larger than its initial mass, $M_2(0).$ Such a star is called
a blue straggler, in analogy to stars in clusters that appear to be more massive
than the cluster turn-off. 

In this section we consider the case in which
$M_c$ is smaller than roughly $0.25\, M_\odot,$ because this
applies to KOI-81 and KOI-74 .
(We include more massive
cores in the calculations described in 
sections 3 and 4.) The primary is not a fully evolved giant at the time it fills
its Roche lobe. 
Its mass will be close in value to $M_1(0)$, and its radius can be expressed
as follows.
\begin{equation}
R_d=0.85\, M_\ast^{0.85}+\frac{3700\, M_c^4}{1+M_c^3+1.75\, M_c^4}
\end{equation}
For an isolated star, the value of $M_\ast$ is simply the star's
initial mass. In the next paragraph we discuss its likely value
in systems that undergo two-phase mass transfer.
At the time when the primary first fills its Roche lobe, $q=M_1/M_2>1.$
If $q$ is larger than a critical value $\eta,$ 
a common envelope
will form
and the core of the primary will spiral closer to the
main-sequence companion. 
The final separation, $a_f$ can be expressed in terms of its initial
separation, $a_i,$ the value of $M_c,$ 
the stellar masses $M_1$ and $M_2$ at the time of 
Roche-lobe filling, 
and an efficiency parameter $\alpha$.\footnote{See, e.g., Webbink (2008) for
details. Here we use $\alpha$ to parameterize the both the  binding energy
of the primary's envelope and the  efficiency of ejecting the envelope
from the system. High values of $\alpha$ correspond to more efficient
ejection and larger final orbits. An alternative approach which 
uses angular momentum in place of energy considerations can
also be applied (Nelemans \& Tout 2005). 
For the purposes of this paper we require
only a formulation that parameterizes the resizing of the semimajor axis.}
\begin{equation}   
a_f=a_i\, \Bigg(\frac{M_c}{M_1}\Bigg)\,   
          \Bigg[1 + 
\Bigg(\frac{2}{\alpha \, f(q)}\Bigg)\, 
\Bigg(\frac{M_1-M_c}{M_2}\Bigg)\Bigg]^{-1} 
\end{equation}   
$f(q)$ represents the ratio between the radius of the
donor and the orbital separation (Eggleton 1983). 
While the companion's mass may not change
significantly during the common envelope phase, it may
have increased during the interval prior to it. This is because
mass transfer may have begun in the form of a gravitationally focused wind
as the primary expanded to fill its Roche lobe. 

There is no single value of the critical mass ratio $\eta$. 
This is because mass transfer
can be stabilized by a combination of factors that each
assume values appropriate to a specific binary. These factors
include
the magnitude
of winds ejected from the system, the specific angular momentum
carried by winds, the role of radiation, and the 
donor's adiabatic index.
When a common envelope is avoided, stars with modest cores will donate
mass during two phases. During phase 1, mass transfer 
occurs at a rate determined by the thermal time scale of the
donor, as it attempts to adjust to mass loss. Once the masses have equalized,
the donor has a chance to reestablish thermal
equilibrium.
The donor star may continue to lose mass, but as it comes into equilibrium
with a smaller mass than its initial mass, and with a core that is still
modest (less than roughly $0.25\, M_\odot$), it begins to shrink
into its Roche lobe. This leads 
 to a quiescent interval. $M_\ast$ is the mass of the donor 
during this time of quiescence. The mass of the accretor
is approximately equal to $M_\ast$ as well, although the accretor
may be slightly more massive than the donor at this point.  
Thus, during phase 1, the donor has lost an amount of mass $M_1(0)-M_{\ast},$
and the accretor has gained $M_{\ast}-M_2(2).$ 
Generally, $M_{\ast}-M_2(2) < M_1(0)-M_{\ast},$ since the thermal
time scales of the two stars are different.  

The continued growth of
the core of the first star, 
and the accompanying stellar expansion, perhaps combined with
magnetic braking, eventually cause the donor to fill its
Roche lobe again. Mass transfer during the second phase
of mass transfer is stable and ends when the donor's envelope is depleted.
Depending on the time elapsed since the start of phase 1, 
the donor's core will generally have grown by an amount $\delta.$

\subsection{Gradual Erosion: KOI-81 and possibly KOI-74}

The first question to answer for KOI-81 and KOI-74 
is whether 
each could be an end state of stable mass transfer. If 
the answer is ``yes'', the donor star would have been filling its Roche lobe
until the time mass transfer ceased, and the orbital period is: 
\begin{equation}
P_{orb}=
 0.372\, {\rm days}
 \Bigg[\frac{R_d(M_\ast, M_c)}{R_\odot}\Bigg]^{\frac{3}{2}}
 \Bigg[\frac{M_\odot}{M_c}\Bigg]^{\frac{3}{2}}. 
\end{equation}
Given a measured value of the orbital period, Equation (1) and
Equation (3)    
 combine to produce a set of possible values for $M_\ast.$ and $M_c.$
The value of $M_c$ should correspond to the present-day mass of the core
observed today, and it can be measured through radial velocity or
other methods (R2010, van~Kerkwijk et al.\, 2010 [vK2010]). 
The value of $M_\ast$ represents the mass of the primary, generally as it
was during the quiescent interval between phase 1 and phase 2.

The black points in the top two panels of Figure~1 show the results
for KOI-81 and KOI-74.
The mass of KOI-81b is constrained to lie in a very narrow 
range: $0.215-0.219\, M_\odot.$ 
(Note however that the systematic uncertainty associated with 
using the functional forms in Equation (1) is larger than this formal range
indicates.) 
The mass $M_\ast$ 
ranges from roughly $1.6\, M_\odot$ and $2.4\, M_\odot.$  
R2019 interpreted light curve features in terms of tidal effects and
found the mass of KOI-81b to be $0.212\pm0.031\, M_\odot.$
This is consistent with our mass transfer model.  
The R2010 value is shown as a cyan triangle centered on a cyan line
which delineates the uncertainty limits. 
vK2010
used Doppler boosting, and measured the mass to be $0.3\, M_\odot.$
vK2010 also used the model of steady mass transfer 
 to predict the mass of KOI-81b, and 
found a value of $0.25\, M_\odot.$ Their model estimate was marginally
higher than ours because they used an expression for the orbital period that
did not have any dependence on the donor mass (Rappaport et al.\, 1995).
Such a form is expected to work best for core masses higher than the estimated
mass of KOI-81b. Note, however, that both 
v2010's expression and Equation (3) are approximate.
If {\it Kepler} identifies many similar systems and
mass measurements are possible for a significant subset,
we will learn how to best model the radii of stars with low-mass cores.   

There is a benefit in utilizing the significant contribution of the donor mass
to the radius by invoking Equation (1). Specifically, the measured value
of the orbital period constrains a combination of both $M_\ast$ and $M_c.$ This
provides input for binary evolution calculations which can
identify all possible initial states for the binary we observe today.
We start by computing the amount of mass lost by the donor during phase 2: 
$\Delta=M_\ast-M_c$; and the fraction of this mass accreted by the 
blue straggler: $\beta=(M-M_\ast)/\Delta$, where $M$ is the present-day
mass of the monitored star. We can then derive the
initial state by evolving backwards in time. 
We find the range of possible evolutionary paths by 
sampling a range of values for (1)~$\eta$ and for (2)~$\beta$ during 
phase~1.\footnote{$\beta$ during phase 1 is generally smaller than during phase 2 
because there is
a mismatch between the thermal time scales of the primary and secondary.}   
Typical results are shown in the bottom panels of Figure 1.
The initial mass of the primary can be as high as $\sim 3\, M_\odot,$ or as low as 
$\sim 1.7\, M_\odot.$ These differences have detectable consequences, in that
more mass is ejected from the binary when $M_1(0)$ is high.
Observations can therefore 
test the predictions of each evolutionary channel, for example
the white dwarf's age
and the amount of mass ejected from the binary as a function of time. 

The results of KOI-74 are shown in the panels on the right hand side of Figure 1.
Using our model, we find that the mass of KOI-74b lies in the 
range: $0.142-0.153\, M_\odot.$ This is marginally consistent with the
results of R2010, based on tidal effects ($0.111^{+0.034}_{-0.038}$).
On the other hand, our estimate is smaller  
than 
vK2010's Doppler-boost measurements ($0.22\pm{0.03}\, M_\odot$), 
and also lower than
 predicted by their period/core-mass relationship ($0.20\pm{0.03}\, M_\odot$).
(Note that when the core mass is smaller, the main-sequence contribution to the
donor's radius is more significant.) The same general features of the range of
initial stars that we discussed for KOI-81 also apply to KOI-74, but the lower mass
of today's monitored star is consistent with lower initial masses for 
the components of the binary.

\begin{figure*}
\begin{center}
\psfig{file=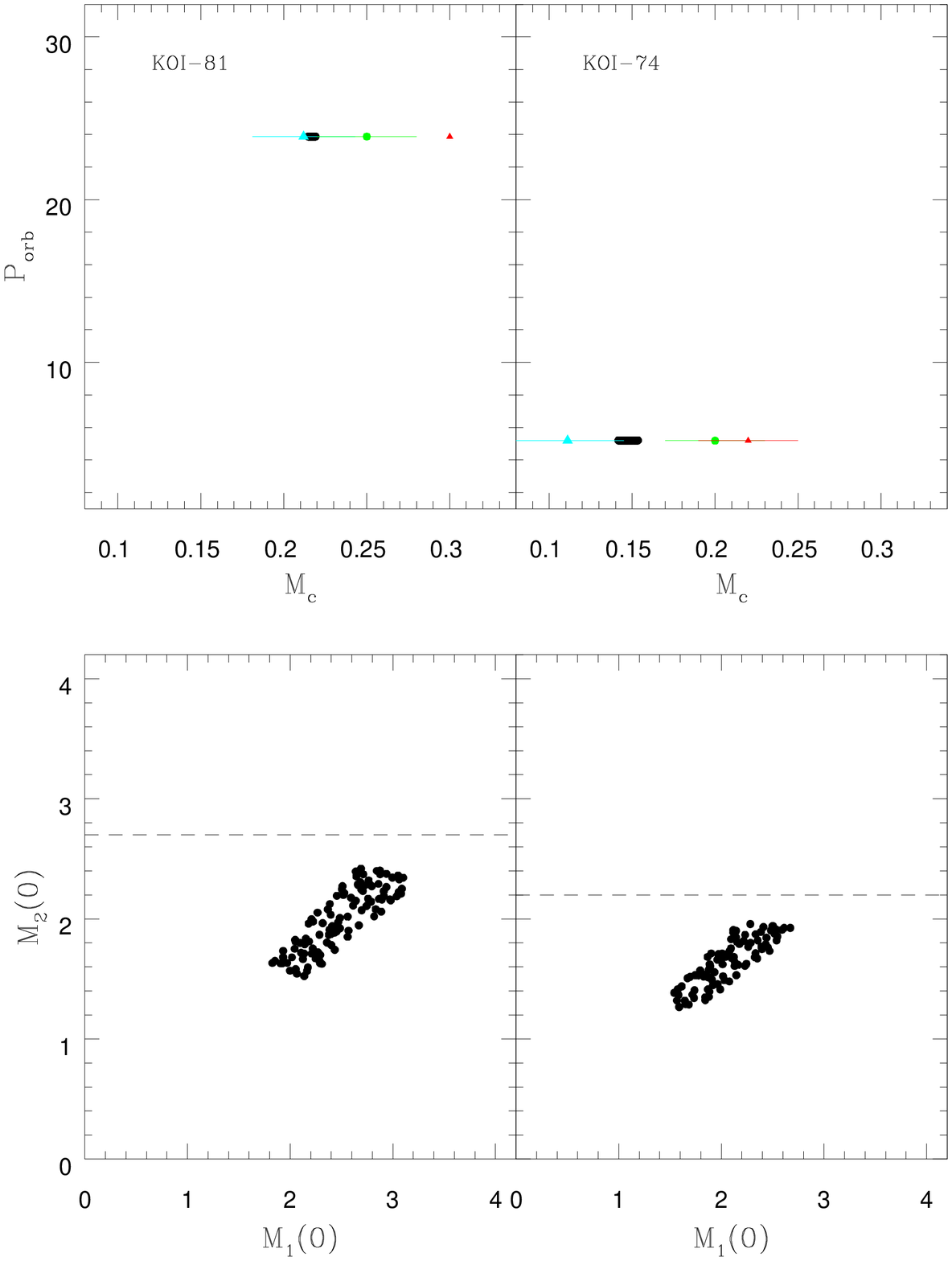,
height=6.5in,width=5.5in,angle=0}
\caption{
Models for the evolution of KOI-74 and KOI-81 that invoke stable mass transfer.
{\bf Top panels:} Orbital period versus the mass of the
white
dwarf.
Black dots show the full range of predictions in this paper. Triangles 
with uncertainty limits show the mass measurements of R2010 (cyan) and 
vK2010 (red). Green points and uncertainty limits show the mass-transfer model
predictions of vK2010.  Note that, in addition to the lower limits shown, 
derived for a tidal-perturbation model, R2010 provide lower limits in the
brown dwarf range; these would not be consistent with the formation of
the systems through stable mass transfer.    
{\bf Bottom panels:}  Initial mass $M_2(0)$ of the secondary versus the
initial mass $M_1(0)$ of the primary for the steady mass transfer
model used  in this paper (\S 2). 
These are the initial states consistent with the
present-day binarity parameters of KOI-74 and KOI-81.
If the correct models are those with higher mass, the systems we observe today
are younger and have ejected more mass.
}
\end{center}
\end{figure*}

\subsection{Disruption: KOI-74?}

If the masses of KOI-74b and KOI-81b are in the range $0.1-0.3\, M_\odot,$
then the work we sketched above shows that 
there are binary evolution models in which their progenitors 
filled their Roche lobes until the time their envelopes were totally eroded.
 This strongly supports the
hypothesis that KOI-71 and KOI-81 are end-states of stable mass-transfer.
We note here, however, that {\it Kepler} can also discover binaries in which 
the mass of the core is  
too large for the donor to have been filling its Roche lobe while in
its current orbit. This would be a signature that the binary
was a common-envelope survivor. We could then use the current values of the
orbital separation and masses to constrain the parameters of the 
common envelope evolution [Equation (1)]. 

For KOI-74b and KOI-81b, however, there may be reason to suggest that the actual  
masses are {\sl smaller} than the values associated with stable mass transfer. 
R2010 place lower limits on the masses 
that extend into the brown dwarf regime. In addition, for KOI-74,
there is only marginal overlap between R2010 and the stable mass transfer model
presented in the last section. It is therefore worth seriously considering that
the mass of one or both of these  cores may be in the brown dwarf regime.
Here we use KOI-74 as an illustrative example. We will show that
a value of the core mass smaller than expected from stable mass
transfer may be a signal that there was a common envelope, but that
the orbit was eccentric at the time the 
primary filled its Roche lobe. This is expected in some triple systems.
It therefore seems inevitable that, whatever the history of KOI-74 and KOI-81,
binaries in which the core mass is too low will eventually be found, and the
scenario we present below will apply. 
 
First we note that, even if the core mass of KOI-74
is small, formal solutions to the stable-mass transfer equations
exist, predicting
$0.005\, M_\odot<M_c<0.018\, M_\odot$ and $1.2\, M_\odot<M_1<2\, M_\odot$ at the
time of Roche-lobe filling.
These solutions are not viable, however, because
the equilibrium radius of a low-core-mass primary would
decline as mass was donated in a steady, stable manner. 
The second term in Equation (1)
would contribute little, and, although the first term might not
apply exactly throughout the evolution, 
the trend of smaller radius with decreasing mass would
be followed during phase 2.
The orbital period would have to be much smaller
than $5.2$~days.

If, therefore, KOI-74b is a low-mass stellar core,
the binary 
likely passed through a common envelope phase. 
Using Equation (2) we 
find that $a_i > M_1/M_c.$ Furthermore, dynamical instability
requires $M_1$ be larger than roughly  $(1.3-1.5) \times (2.2\, M_\odot).$ 
This predicts an initial separation so large that
the primary 
could not have filled its Roche lobe.

This may indicate that the initial orbit was highly eccentric and that
a dynamical instability was triggered during periastron.  
During the common
envelope, the eccentricity was 
eliminated while the semimajor axis became
smaller. 
 Such a high eccentricity is not generally expected in a binary in which
one companion is poised to fill its \rl , because tidal effects will tend 
to circularize the orbit.  If, however, KOI-74 is a triple, the 
Lidov-Kozai (Lidov 1961; Kozai 1962) mechanism
could have produced the extreme eccentricity that led to the
disruption of the progenitor of today's low-mass core. 
Perets \& Fabrycky (2009) have considered the generation of blue stragglers
via this mechanism. They focused on mergers, but the same general idea
could apply to systems like KOI-74. We therefore propose that mass transfer
occurred prior to the approach that triggered the common envelope.
During that earlier epoch, the progenitor of the hot core made
close approaches to the star we monitor today. In a process 
analogous to what happens in high-mass
x-ray binaries, mass was transferred during these close approaches. 
Eventually, under the influence of a third star,
 the periastron distance became small enough to trigger a dynamical instability.  

We used Equation 1 to compute the initial value, $a_i$, of the 
binary's semimajor axis at the time the primary filled its Roche lobe.
The secondary mass was considered
to be the  mass observed today, 
and we set $M_c$ to $0.03\, M_\odot,$ 
$a_f=16\, R_\odot,$ and assumed efficient envelope ejection ($\alpha=100$),
so as to compute the minimum value of $a_i.$ 
We found that the eccentricity was required to be large, generally
 $> 0.997.$ 
Thus, this scenario predicts the presence of a third body in the system.
Its separation from the $2.2\, M_\odot$ star {\it Kepler} is
monitoring today would likely be greater than
a few hundred AU (See Figure 2), so that this 
star third is potentially detectable today.
As shown in Fabrycky \& Tremaine (2007) and 
references therein, triple stars are 
common, and a large fraction of close binaries are known to have
another companion in a wide orbit. (See also 
Eggleton \& Kisseleva-Eggleton 2001.) 
This triple-star scenario therefore seems plausible. The one feature that
seems unlikely is the extreme value of the eccentricity required. 

Finally, we note another, less likely possibility.
Eggleton (2002) discusses systems in which the envelope was so
weakly bound to the primary that dynamical instability did not produce
substantial spiral-in. This situation does not seem likely to apply to 
KOI-74.  

The fact that different values of the mass of today's small-radius core 
are associated with different evolutionary scenarios points to the
importance of reliable mass measurements for both KOI-74 and KOI-81.
While it is intriguing that the well-sampled {\it Kepler} light
curves have allowed both Doppler boosting and tidal methods to be
employed, a radial velocity measurement is required. Preliminary indications
are that the 
rotational velocities of both
KOI-74a  and KOI-81a are high, perhaps in the range of
$\sim 300$~km~s$^{-1}$ (Latham 2010). While this may complicate 
radial velocity measurements, it could also be a clue that 
each has accreted matter, thereby supporting the mass transfer model.   
High rotational velocities are, in fact, 
often associated with blue stragglers, presumable generated through
mass transfer (see, e.g., De~Marco et al.\, 2005 and Mathieu \& Geller 2009). 
While, therefore, the high rates of rotation
may make mass measurements more difficult, it is also possible
that they are providing a signal that mass transfer did occur.

\begin{figure*}
\begin{center}
\psfig{file=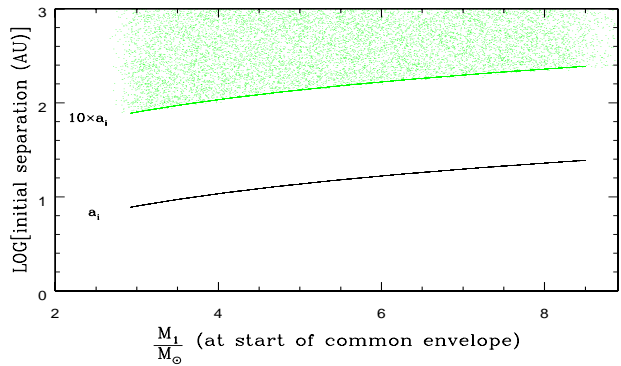,
height=4.0in,width=3.5in,angle=0}
\vspace{-1.9 true in}
\caption{
Common envelope evolution for KOI-74. Lower (black) curve
shows $a_i$ versus $M_1(0)$ 
for efficient envelope ejection.
If the mass of the hot core detected by R2010 is in the brown-dwarf range,
then the orbit may have been eccentric, and the massive donor star 
may have experienced a dynamical-time-scale instability at periastron. The high eccentricity would
likely have 
been induced  
by interactions with a third star.
Plotted in green are the possible  orbital separations of the third star from
the star monitored today by {\it Kepler.}
}
\end{center}
\end{figure*}
\vspace{-0.4 true in}
\begin{figure*}
\vspace{-0.5 true in}
\begin{center}
\psfig{file=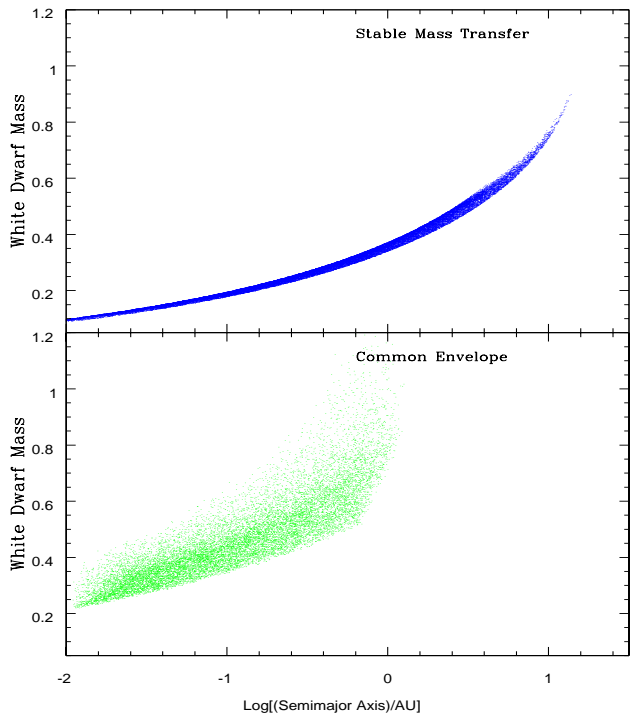,
height=4.0in,width=3.5in,angle=0}
\vspace{-.10 true in}
\caption{
$M_c$ versus log of the current orbital separation.
Binaries that have experienced
stable  mass transfer
(top panel);
common envelope evolution
(bottom panel).
}
\end{center}
\end{figure*}
 

\section{Probability}

The advent of a new observing capability has almost always led
to unanticipated discoveries. The two
binaries, KOI-74 and KOI-81, may be examples.
Nevertheless,
we may question whether the evolutionary models
presented above are expected to occur commonly enough
 that {\it Kepler} should have been
able to discover these interesting binary systems by monitoring
only $\sim 150,000$~stars.
We therefore conducted a first-principles study of binaries in a stellar
population to predict the numbers of transiting systems we expect
to be comprised of a main sequence star orbited by a
white dwarf
that has emerged from an episode of mass transfer. 
Although these calculations
were suggested by the discoveries of KOI-74 and KOI-81,
they do not rely on the interpretation of these 
systems. We note that the evolution of interacting binaries consisting of
a main sequence accretor and a subgiant or giant donor is complex and
involves a wide range of physical processes. There are uncertainties,
including the results of common envelope evolution, the fraction
of incoming matter that can be retained by the accretor, and the
angular momentum evolution of the system. Nevertheless, by parameterizing
the effects of these processes we are able to derive a robust 
conclusion.   

The {\it Kepler} team selected targets from roughly half a million
stars in its field brighter than 16th magnitude.
More than $90\%$ of the targets were selected based on signal-to-noise
considerations that suggested the possibility of detecting terrestrial-size
planets. While a small fraction of the targets are selected
to pursue a range of other science opportunities, including. e.g.,
 eclipsing binaries and high-proper-motion stars,
the majority of the targets ($\sim 90,000$)
are G-type stars on or near the main sequence (Batalha et al.\, 2010).
The presence of KOI-74 and KOI-81 illustrates the presence of
more massive main-sequence stars as well.

In the calculations described below, we compute the fraction of monitored 
stars transited by \wdf s per year. The number   
of systems in which \tr s can be \det ed is the product of this fraction
and the number of monitored stars with high enough signal-to-noise that the
\tr s of \wdf s can be \det ed. Massive \wdf s have radii comparable to
the radius of the Earth. Transits of many of the \kep\ target stars
by objects of this size should be detectable. The \wdf\ radius increases
with decreasing mass (see. e.g., Parsons et al.\, 2010), 
so the less massive \wdf s which are expected to
be common among \m-t\ products should also produce \det able \tr s when they
pass in front of $90\%$ of the\kep\ targets. White dwarfs that have not yet 
had a chance to cool are even larger. We therefore define a quantity
${\cal N}_{mon}^{detect},$ the number of monitored stars against which
\wdf\ \tr s can be \det ed.\footnote{Although ${\cal N}_{mon}^{detect}$
is a function of the white dwarf mass and age, this is a relatively
small effect. The value to which we normalize can therefore be viewed
as a sort of average.}
The value of ${\cal N}_{mon}^{detect}$ is roughly 
$0.9 \times 150,000 = 135,000.$ As we will discuss in \S 4, the decline 
in received light produced by a transit can be balanced by the gravitational
lensing
produced by the \wdf . While this may make it difficult or impossible
to \det\ \tr s in some systems, lensing produces strong, distinctive
signatures, including ati-transits, in others.  

\subsection{Calculations}

We can derive a rough estimate of the fraction of {\it Kepler}
 targets that might be orbited by transiting white dwarfs by
generating a population of stars, including both
singles and binaries, and then
considering the evolution of each system.  We identify those that will
become main sequence stars that are both (1)~in a
range of masses likely to be monitored by
{\it Kepler}, and (2)~orbited by white dwarfs. We then compute the
probability that the white dwarf will transit during the time of
{\it Kepler} monitoring. For short
orbital periods, the probability is $R_\ast/a$, where $R_\ast$ is the
radius of the monitored star and $a$ is the final orbital
separation. For orbital periods longer than
$\tau_{monitor}$, the duration of monitoring, 
 we multiply by a factor of
$\tau_{monitor}/P_{orb}.$

We assume that $40\%$ of all stellar systems are singles, and that
the remainder are of higher multiplicity.  We use a Miller-Scalo
IMF (Miller \& Scalo 1979)
to select the masses of all single stars, and to select the
primary mass for all multiples.
This has the effect of
favoring low-mass stars, even though we generate the
full mass spectrum.
The only multiple systems we explicitly consider are binaries.
We select the secondary mass as
follows: $M_2=M_1\times u^{0.8},$ where $u$ is a random number
chosen from a uniform distribution in the range $0-1.$
We then generate orbital separations, selecting uniformly in log space
from separations as small as three times the radius of the
primary, and as large as $10^4$~AU.

Primary stars will fill their Roche lobes at some
point in their evolution
   if the orbital separation lies within roughly an AU. This range
of separations encompasses a large fraction of all binaries.
Roche-lobe filling will lead either to a common envelope phase or
to stable mass transfer. While a large fraction of common envelopes
may end in mergers, those binaries that
survive will have small orbital separations and therefore high
probability of yielding transits.
When mass transfer is stable, the final
separation
may be somewhat larger than the initial separation, but most orbits are
smaller than a few AU (see Figure 3).

The paragraph above describes the key reason why \tr ing \wdf s 
must be common: they are produced by a large fraction of all
primordial binaries.   
We used a simple model to design a code that would allow us to
quantify this and to test
a wide range of input assumptions. 
We conducted $13$ simulations.
Each was characterized by (a)~the value of $\alpha,$ (b)~the value of
$\tau_{monitor},$ (c)~the range of masses of stars monitored by 
{\it Kepler,} (d)~the rule used to compute the final core mass for
systems experiencing stable mass transfer, (e)~the ages of the 
stars monitored by {\it Kepler.}  
In this way, we were able to take into consideration the uncertainties
in the input physics and to identify those results that are robust. 
The key question is: what fraction of all monitored systems
are binaries in which a white dwarf is expected to transit a 
main-sequence star? 

To compute the number of monitored systems we 
counted members of the following sets of stars, considering
only those in the appropriate range of masses: 
(1)~all isolated stars;
(2)~stars that are the results of
mergers; 
(3)~stars with dwarf stellar companions having luminosity less than $0.001$
their own luminosity\footnote{Depending on the orbital inclination these might
provide transits that appear to be planetary. They could also exhibit 
deep eclipses like those seen in KOI-74 and KOI-81.}; 
(4)~stars   
in binaries so wide that the individual stars 
could be resolvable\footnote{Because
we do not attempt to generate the 3-D spatial distribution of stars, it is
not clear which binaries will actually be resolvable. We assumed that all
binaries with orbital separation greater than $10^{3.5}$~AU were potentially
resolvable. By adding to the number of apparently isolated stars, we are
lowering the fraction of systems with white dwarfs orbiting main sequence stars.
This produces a more conservative estimate.}; and 
(5)~stars orbited by white dwarfs. This last category includes
all white-dwarf/main-sequence pairs, regardless of the formation history
of the white dwarf.

Figure 3 shows the result of one such simulation. In this particular 
simulation
we generated $4$ million stellar systems, and 
used $\alpha=10,$ and $\tau_{monitor}=1$~year. 
We assumed that {\it Kepler} is monitoring main-sequence stars with
masses in the range $0.9-2.9\, M_\odot.$ We assumed that, for systems
undergoing stable mass transfer, the final value of the core mass
would lie within $10\%$ of its initial value ($\delta < 0.1\, M_c$). 
There were approximately $370,000$
stars with masses within the selected range 
that would appear to be single stars.
We then identified all
 binaries in which the primary star fills its Roche lobe. Those with
$q> 1.5$ were assumed to undergo common envelope evolution. For smaller values
of $q$, we assumed stable mass transfer and selected the final value of the 
primary's core mass.
The numbers of binaries with the present-day primary having a mass in
the monitored range and (1) surviving the common envelope, were $\sim 16,000;$
(2) having ended stable mass transfer, were $\sim 35,000.$ 
Most of these will not
produce transits. Taking the transit probability into account, we find that
$3.6\times 10^{-3}$ of the monitored stars should exhibit at least
one transit per year, in which the transiting object is a white
dwarf that survived a common envelope. Many of these will exhibit multiple
transits per year.  
The comparable fraction of systems with with transiting white dwarfs that
emerged from stable mass transfer is $5.0\times 10^{-3}$.

Table~1 shows a representative sample of the results. In this section
we focus on the totals listed in column 6. These totals are normalized
to the case in which \tr s can be \det ed in $135,000$ monitored stars 
The first two rows correspond to a single simulation, labeled ``1''.
Simulation 1 is the one used
to produce Figure~3. ($\alpha=10, \delta < 0.1,$ the only criterion
imposed on the  \kep\ target stars is that their masses are 
between $0.9\, M_\odot$ and $2.9\, M_\odot.$)  
If this case applies, more than $1100$ transiting systems would be
discovered by \kep . 
In the next 4 rows, also labeled ``1'',  we alter
only the value of $\alpha,$ the \ce\ ejection efficiency. 
We find that, as the ejection
efficiency declines, significantly more mergers occur.
 For example
when $\alpha$ declines from $10$
to unity, the numbers of mergers increase by $50\%$ and
the common-envelope survivors are found in closer orbits than
those shown in Figure 3. The transit
probability declines by only $15\%$, however, because transits are 
more likely for closer orbits.  
The value of $\alpha$ must decline to $\sim 0.1$ in order for \ce\ 
survivors to produce fewer than $100$ transits per year. 
Thus, the result that a large number of \wdf s should
produce \tr s is robust with respect to changes in
the common-envelope efficiency factor.   
In any realistic population, the efficiency of envelope ejection
will be different for different systems. To determine the
numbers of transits expected from \ce\ survivors, we should therefore
average over a range of efficiencies.

We also tested the effect of changing the rule used to compute
the final core mass for the donor stars that contribute mass
during stable mass transfer. 
We used three prescriptions: 
(1)~the maximum
value of $\delta$ is $10\%$; (2)~the maximum
value of $\delta$ is $20\%$; (3)~the value of $\delta$ can be as large as 
half the difference between the white dwarf mass expected for a star
with initial mass $M_1(0)$ and the core mass $M_c$ at the start of mass
transfer. 
Changing these rules doesn't change the numbers of stable \m-t\ binaries,
but it does affect 
the 
transit probability; 
the larger the
 core mass at the time of depletion, the wider the orbit, and the smaller
the transit probability. 
With prescription 2, the numbers of transits by white dwarfs that are
the end states of stable mass transfer
decreased by $\sim 13\%.$  With prescription 3,
the numbers of transits by the end states of stable mass transfer
decreased by roughly $70\%.$ 

No single rule for selecting a final core mass 
can apply to all binaries. Once mass transfer begins, the donor's envelope
is depleted through mass loss, and the envelope is also contributing
mass to the growing stellar core. The final core mass depends on 
the time scales for these two competing processes. These time scales
depend on the state of the system and on the evolution of the
orbital angular momentum, for example on the amount of angular momentum  
carried off in winds. 
For donors that fill their Roche lobes when their core masses are
small ($\sim 0.1-0.2\, M_\odot$), mass transfer may deplete the envelope
at a faster rate than the nuclear evolution of the core.
Values of $\delta$ may be modest. For donors that fill their
Roche lobes as more evolved stars, the evolution of the core
is proceeding more rapidly. The rate of mass loss will also typically be high,
however, and can be increased if winds carry significant angular momentum.   
To incorporate these effects, we could carry out a detailed evolution for
each system. The uncertainty would still be large, because the
physical effects that determine the binary evolution are not yet
well-enough understood. To determine the numbers of transits
expected by white dwarfs that are the remnants of Roche-lobe-filling
donor stars, 
 we have therefore chosen to average over the 
three prescriptions described above
 for the final mass of the core. 
This averaging process corresponds to considering a range of
mass-transfer rates as well as evolutionary states for the donor.

The age of the monitored systems can also influence the results.
In particular, the primary needs to have had enough time to evolve
and produce at least a modest core. Yet the star that was originally
less massive should still be on the \ms, even after gaining some
mass from its companion. 
To take this into account, in simulations labeled ``2'', 
we considered a constant rate of
star formation over an interval 
of $12$~billion years.
We 
randomly generated the time at which each stellar system was formed, and
considered further  
only single stars which would not have had time to evolve during the
interval from their formation until the present day. We considered 
only binaries in which the primary star had time to evolve, but
the secondary is still on the main sequence.       
In this case the fraction of transiting systems derived from 
common envelope evolution (stable mass transfer evolution) was
$4.3\times 10^{-3}$ ($2.0\times 10^{-3}$). 

We conducted additional simulations, not shown in the table, because
the results did not change significantly. 
For example, changing $\tau_{monitor}$ from $1$~year
to $3.5$~years changes the results by only a few percent, 
since most of the transits
occurred in binaries with orbital periods smaller than a year. (Multiple
transits of the same monitored star will, however,
 increase the reliability of the detection.)
As another example, monitoring stars in a more limited 
mass range ($0.9-1.9\, M_\odot$, or $1.9-2.9\, M_\odot$) changes the results
by only $\sim 10\%.$ 

\subsection{Results}
It is useful to consider the contribution of common-envelope
 survivors and stable-\m-t\
binaries separately.  
\begin{equation}   
N_{transit}=N_{transit}^{ce} + N_{transit}^{mt} 
\end{equation}   

To estimate the value of each term, we average over the results for the 
simulations described above. First, we note that simulation 1 is more
appropriate for an old population in which the white dwarfs would have
long-ago formed from  
 the primary stars in
binaries with secondaries 
as massive as the \kep\ targets. 
Simulation 2 applies to intermediate-age populations. 
We average the results of these 
two simulations. In each case we average over the common envelope
efficiency factors and also over the prescription for the value of
$\delta.$  

\begin{equation}  
N_{transit}^{ce}=\frac{{\cal N}_{monitored}^{detect}}{135,000}\, \Bigg[395 \pm 32\Bigg]   
\end{equation}   
and
\begin{equation}  
N_{transit}^{mt}=\frac{{\cal N}_{monitored}^{detect}}{135,000}\, \Bigg[338 \pm 143\Bigg]   
\end{equation}   

Including stellar systems with higher mutiplicity would increase the 
numbers of systems with white dwarfs in close orbits.

\subsection{Science Returns}

The calculations described
above suggest that {\it Kepler} may identify transits in roughly a thousand  
mass-transfer end states.  Each case provides a unique
way to measure the white dwarf radius, luminosity and temperature, and
even to
explore its atmosphere. This is all
in addition to
the standard ``bag of tricks'' that can be applied to any white dwarf in
a binary. The ability to conduct a large number of such studies in a uniform way
will determine the properties of the white dwarfs in terms of the
mass lost history of their progenitors.

With regard to mass loss histories,
{\it Kepler} studies will
establish the relative frequency of common envelope evolution and
provide useful data points for determining the best way to compute common envelope
evolution  
in terms of the properties of the binary at the time the dynamical
instability was triggered. The range of systems producing
stable mass transfer will also be better understood, including the
all-important fraction, $\beta,$ of material that can be retained by
a main-sequence accretor. Most of the {\it Kepler} stars are
field stars. Studying the {\it Kepler} blue stragglers will
provide empirical insight into the contribution of ordinary
binary evolution to the creation of blue stragglers.
This will allow the contribution of stellar interactions
in  clusters to be better understood, since both primordial binaries and
binaries formed or altered through interactions can produce blue stragglers.

Finally, the white-dwarf/main-sequence binaries studied by {\it Kepler}
will experience a second phase of interaction when the main-sequence star
of today evolves and begins to transfer matter to the white dwarf.
These systems will be novae, some may become double-degenerates that
eventually merge, and some 
may even experience Type Ia supernovae, or
accretion-induced collapse, 
through either the single-degenerate or double-degenerate
channel. 
{\it Kepler} will identify
a set of these systems that can be studied in a unique way.

\section{Mass Measurements and Lensing: White Dwarf or Terrestrial Planet?}
The primary goal of the {\it Kepler} mission is to search for terrestrial 
planets orbiting sun-like stars, especially in the zone of habitability.
We do not yet know how common such planets are. The results described in
\S 3 establish, however,
 that orbiting white dwarfs are well represented in the 
{\it Kepler} data.  White dwarfs have radii similar to planetary radii and,
as Figure~3 shows, many are likely to orbit in their host star's zone
of habitability.  
It is therefore important to be able to distinguish between planets and
white dwarfs. 

The discoveries of KO-74b and KOI-81b illustrate that it is
possible to identify transiting objects that are candidate white dwarfs.
Yet, even though both systems are likely to be products
of mass transfer, the natures of the hot compact objects and
their evolutionary histories are not yet definitely established.
Furthermore, not all white dwarfs will be hot and luminous enough
to produce distinctive eclipses. While dwarfs cool with age, and 
``ultracool'' white dwarfs with temperatures below $5000$~K have been
discovered (Vidrih et al.\, 2007; Harris et al.\, 2008). 

The key property that is different for planets and white dwarfs,
irrespective of age, is the mass. Mass measurements are therefore
of central importance, ideally through radial velocity measurements.
Interestingly enough, this may be difficult to do for some
white-dwarf blue straggler systems, because high rates of stellar
rotation can result from mass transfer and complicate the
measurements. On the other hand, high rotational velocities may
be a signature that supports the mass transfer scenario. 
As mentioned in \S 2, there are preliminary indications that
the rotation rates of KOI-74a and KOI-81a are high
(Latham 2010).

Even if high rates of rotation are rare, the large numbers of interesting systems 
discovered by {\it Kepler} may make it challenging to obtain 
radial velocity measurements for all of them.
It is therefore useful that, 
because of its sensitive photometry, {\it Kepler}
is well-suited to make optimal use of light curve deviations to 
measure mass. 
Both tidal effects and, for the first time,
Doppler boosting have been used to provide mass estimates of 
KOI-74b and KOI-81b. 
If a modest number of systems can be studied through
both light-curve and radial-velocity techniques,   
we may be able to rely more heavily on the former. 
In addition,
when the transiting object is a stellar
remnant, its high density may allow us to detect its action as
a gravitational lens that deflects light from the 
star it orbits. Measuring the lensing effects can provide an independent
estimate of the gravitational mass of the white dwarf.  

The Einstein radius is
\begin{equation}
R_E= 2.97\times 10^9 {\rm cm}\, 
      \Bigg[\frac{M_{WD}}{M_\odot}\, \frac{a}{\rm AU}\Bigg]^{\frac{1}{2}} 
\end{equation}
In the cases of interest to this investigation, 
the Einstein radius is comparable to the radius
of the white dwarf, which is the lens.
Finite-lens effects
must therefore be considered; they diminish the magnification and
make the signatures of lensing more subtle. 
 This is not the only complication,
since the Einstein ring is smaller than the lensed
star.
Finite-source-size and limb darkening also
influence the value of the peak magnification
and the shape of the light curve. 
A set of instructive
examples that take these effects into account can be
found in the pioneering investigation of Sahu \& Gilliland (2003),
which studied the detectability of binary self-lensing during transits,
as observable by
{\it Kepler.}  

While the transiting mass blocks light from the star it orbits,
its action as a lens increases the amount of light we receive.
Sahu \& Gilliland (2003) found that, 
for low-mass white dwarfs in close orbits, the transit signature 
dominates. As the mass and/or orbital separation increases, lensing
begins to play a larger role and can effectively balance the diminuation
of light associated with the transit. The transits of
some white dwarfs will therefore not be detectable,
even though the cancellation is not exact. As the mass and separation
continue to increase, the lensing magnification dominates. During transit,
and for an interval both before and afterward, 
there is a highly significant increase in the received light.
For systems in which the influence of lensing can be photometrically detected,
model fits can provide estimates of the white dwarf's
mass and radius. 

Sahu and Gilliland (2003) showed that {\it Kepler} would be likely to
detect lensing by white
dwarfs, but did not have a detailed model of white-dwarf/main-sequence
binaries and could therefore not predict the rates.
The population calculations of the previous section can be used as
a basis to compute the expected rates. Ideally, the light curves
of each transiting system formed in the population calculation would be  
computed.   
This is a large project, because a variety of other effects must
also be included; in addition, the age of the white dwarf influences its radius.
Here we attempt a simple estimate, with the goal of determining
how frequently signals of lensing in the light curve may eliminate
confusion with planets. We utilize the value of    
$R_E/R_{WD}$ to test the detectability of lensing effects,
using guidelines derived from the work of Sahu \& Gilliland (2003).
Note that, at almost the same time as Sahu \& Gilliland (2003),
 Farmer \& Agol (2003) took an
alternative approach to incorporating the effects of lensing, in the
context of a population synthesis analysis. Although they did not classify
systems by their previous evolution (common envelope versus
stable mass transfer), they did provide numerical results which can be
compared with our totals (see \S 5). 
\begin{table}
 \centering
\caption{Transiting White Dwarfs: Numbers and Lensing Effects}
\vspace{.1 true in}
\label{tlab}
\begin{tabular}{cccccc}\hline
Method/Type  & $\frac{R_E}{R_{WD}}>2$ & $0.8 < \frac{R_E}{R_{WD}}<2$ & $0.4 < \frac{R_E}{R_{WD}}<0.8$ & 
    $\frac{R_E}{R_{WD}}<0.4$ & {\bf Total}\\
\hline
\hline
1:CE ($\alpha=10$)                           & 6.3  & 40.0  & 102.7 & 343.6 & {\bf 492.6}  \\
1:MT ($\delta_{max}=0.1$)                    & 3.4  & 26.6  & 58.5  & 589.3 & {\bf 677.8}  \\
\hline
1:CE($\alpha=3$)                             & 4.3  & 44.0  & 116.5 & 311.6 & {\bf 476.4}  \\ 
1:CE($\alpha=1$)                             & 1.3  & 42.8  & 139.4 & 243.8 & {\bf 427.3}  \\ 
1:CE($\alpha=0.3$)                           & 0.2  & 22.9  & 144.4 & 159.4 & {\bf 326.9}  \\ 
1:CE($\alpha=0.1$)                           & 0.0  &  0.4  &  30.3 &  61.7 & {\bf  92.4}  \\ 
\hline
1:MT ($\delta_{max}=0.2$)                    & 3.1  & 25.5  & 55.2  & 506.1 & {\bf 590.0}  \\
1:MT ($\delta_{max}=\delta_{max}(M_c)$)      & 3.5  & 26.6  & 40.1  & 117.1 & {\bf 187.3}  \\
\hline
\hline
2:CE ($\alpha=10$)                           & 4.1  & 36.4  & 94.9  & 445.4 & {\bf 580.8}  \\
2:MT ($\delta_{max}=0.1$)                    & 1.1  &  9.7  & 21.4  & 239.1 & {\bf 271.3}  \\
\hline
2:MT ($\delta_{max}=0.2$)                    & 1.0  &  9.3  & 20.4  & 204.5 & {\bf 235.2}  \\
2:MT ($\delta_{max}=\delta_{max}(M_c)$)      & 1.2  & 10.8  & 16.5  &  50.2 & {\bf 78.7}  \\
\hline
\hline
\end{tabular}
\end{table}

\begin{figure*}
\begin{center}
\psfig{file=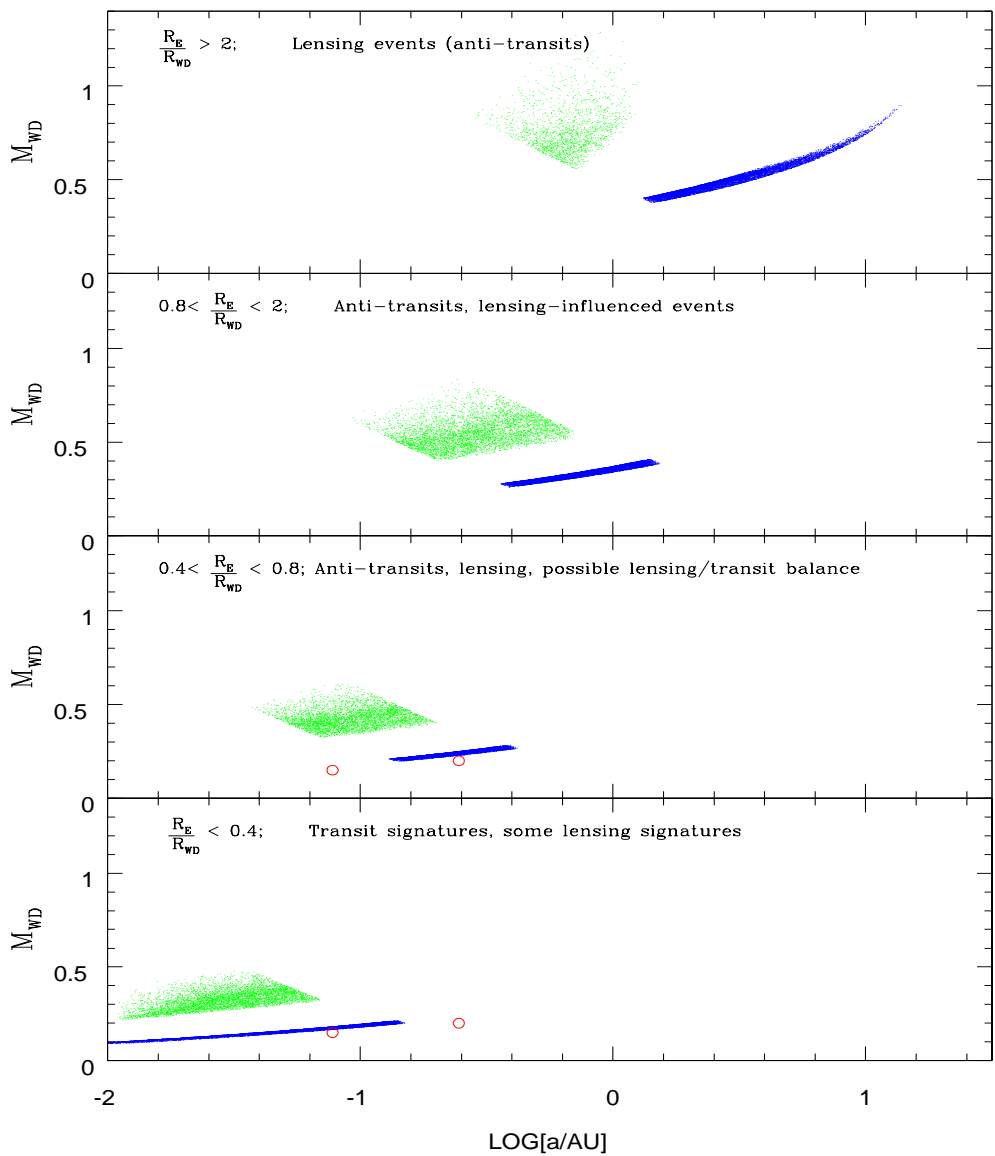,
height=6.5in,width=5.5in,angle=0}
\caption{
$M_c$ versus log of the current orbital separation.
Dark blue points represent binaries that have experienced
stable  mass transfer; light green points represent binaries 
that have experienced a common envelope.  
Points in each panel have values of $R_E/R_{WD}$ in the range shown.   
The two red points shown in the third and fourth panel 
correspond to KOI-74 (on the left) and KOI-81 (on the right and slightly
higher). Both KOI-74b and KOI-81b have larger radii than the
mass/radius relationship for \wdf s would indicate, perhaps indicating that
they have not yet cooled. Each has values of $R_E/R_{WD}$ less than
or equal to $\sim 0.1.$   
}
\end{center}
\end{figure*}

\begin{figure*}
\begin{center}
\psfig{file=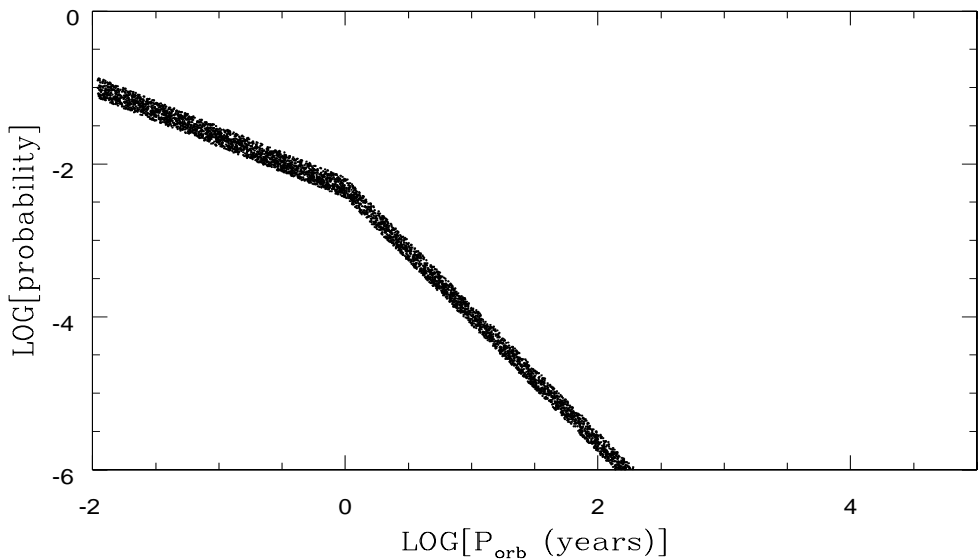,
height=6.5in,width=5.5in,angle=0}
\vspace{-3.2 true in}
\caption{
Low-mass stellar dwarfs may produce transits that mimic planetary transits.
Here we plot the probability of a transit versus the logarithm of the 
orbital period. The bend at one year corresponds to $P_{orb}=\tau_{monitor}.$
For shorter orbital periods, the probability falls as $1/a$; for longer orbital periods
the ratio between $\tau_{monitor}/P_{orb}$ must also be factored in. 
The presence of low-mass dwarfs with luminosities in the 
range of $10^{-4}-10^{-3}$ times the luminosity
of the star monitored by {\it Kepler} may be detectable only because they
produce detectable transits, and are themselves detectably eclipsed.
We kept track of these systems in our simulations. Considering only those with
orbital periods greater than $4$~days, we found that they should produce
transits at $\sim 5\%$ the rate caused by white dwarfs that emerge from common envelopes.
The total contribution is likely to be higher.
}
\end{center}
\end{figure*}

\begin{figure*}
\begin{center}
\psfig{file=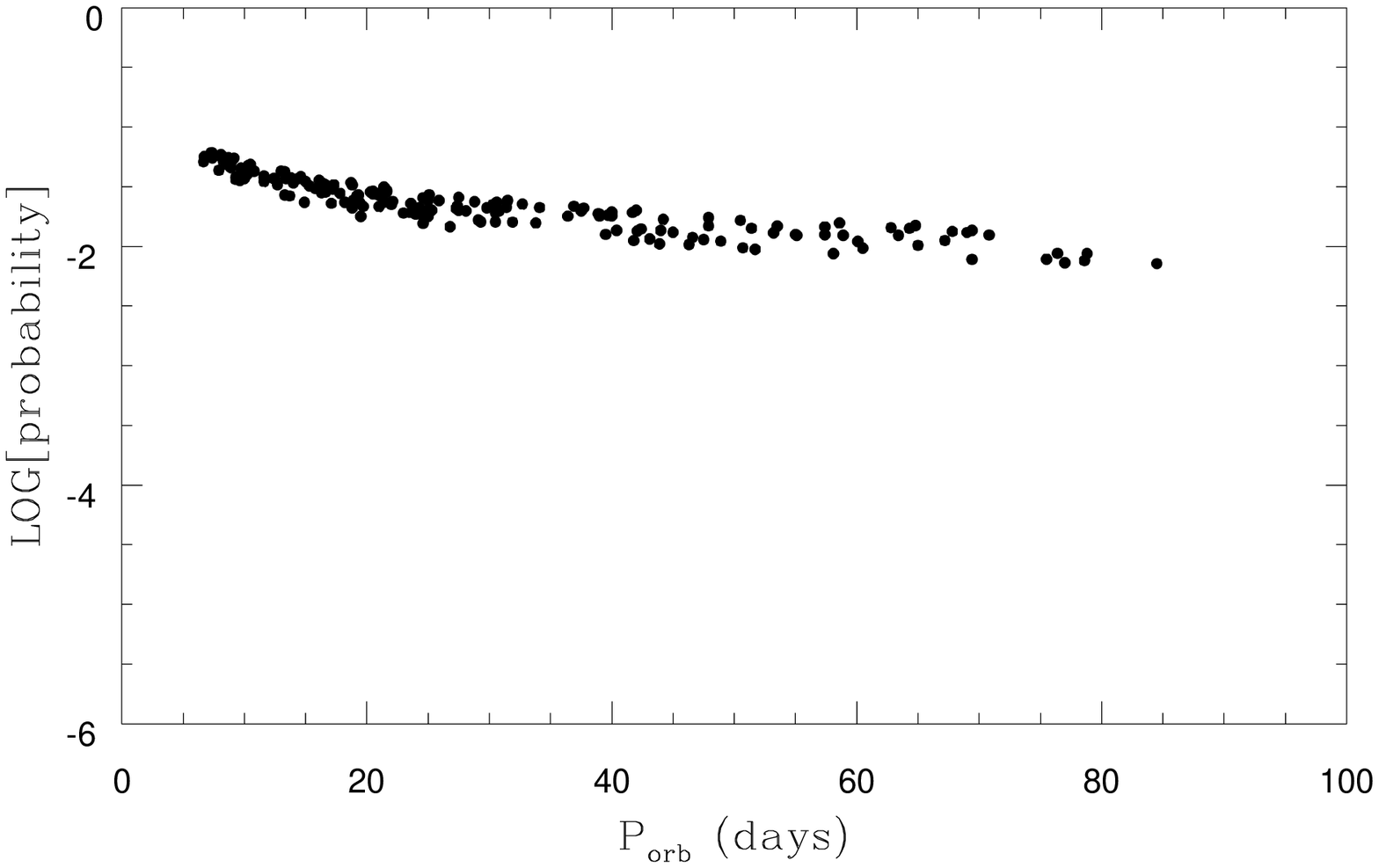,
height=6.5in,width=5.5in,angle=0}
\vspace{-2.8 true in}
\caption{
Log of the probability of detecting a neutron star or black hole 
lensing event versus the orbital period. As with transits, these events
would repeat on a periodic basis as the neutron star or black hole
executes an orbit around the star monitored by {\it Kepler}.
 {\it Kepler} could detect
a handful of events caused by compact objects more massive than
white dwarfs.  
}
\end{center}
\end{figure*}

 We computed the value of $R_E/R_{WD}$ for 
every white dwarf in our simulations,  using 
an empirically-derived approximation to estimate the
radius of a white dwarf of given mass $M_{WD}.$ From Parsons et al.\, 2010
we use: $R_{WD}= [0.014-0.012\, (M_{WD}-0.5)].$   
The results are shown in Figure 4 for the same simulation that
produced Figure 3. 
Note that Figure~4 provides a guide to the characteristics
of binaries in which lensing will be important. It shows, for example,
 that
common envelope systems are more likely to be affected by lensing.
We note, however, that for individual systems, the effects must
be computed separately. This is because the radius of the white
dwarf may larger or smaller; finite source size effects can be important;
and, the range of values of $R_E/R_{WD}$ encompassed in a single
panel correspond to light curves with a range of properties.

The systems shown in the top panel Figure~4 all have
$R_E/R_{WD}>2.$ In these cases we expect the lensing signature to be 
clear and for the gravitational mass to be potentially measurable.  
In this regime, many of the events will look like ``anti-transits''.
For the sake of clarity, we note that we are using the term ``event'' to
refer to the light curve deviation associated with a single passage of the
white dwarf in front of its companion star. Just as with ordinary transits,
these events repeat periodically, and the ability to detect
repetitions increases confidence 
in the physical interpretation. 
Figure 15 of Sahu \& Gilliland (2003) shows the lensing signature for
a white dwarf  of $0.6\, M_\odot$ located $1$~AU form a solar-type star.
The light curve exhibits what we have referred to here as an anti-transit,
with a significance in the {\it Kepler} data greater than $100\, \sigma$.
The significance of some of the events generated by systems in the top panel of Figure 4
will be even higher, and some of these events may therefore be detectable
even from the ground. 

Figure~14 of Sahu \& 
Gilliland (2003) shows that even with  $R_E/R_{WD}= 0.8$, an anti-transit
can be clearly detected. 
The second panel shows systems with $0.8<R_E/R_{WD}<2.$
Some of these events will be anti-transits as well; 
lensing should be recognizable in many.  
Those that are not pure anti-transits are likely to exhibit
 detectable deviations from baseline that exhibit  
significant signatures of lensing. 
From the perspective of studying white dwarfs
and the end states of mass transfer, events from systems in these top two panels
are important, because lensing provides 
a unique way to measure the white dwarf mass.
From the perspective of {\it Kepler's} primary goal of identifying transits by
terrestrial planets, these events are also important, because the degeneracy
with transits by planet-mass lenses is clearly broken.

The third panel shows systems with $0.4<R_E/R_{WD}<0.8$ Some of these may also
produce clear signatures of lensing. Others, particularly 
when the white dwarf is young
and still larger than suggested by the mass/radius relation,  
may produce events that look like
ordinary transits. In addition,
a significant subset of the 
events generated by this group will exhibit some level of
cancellation between diminuation due to transit and magnification
caused by lensing. That is, some systems 
shown in the third panel generate events (combinations of transit and lensing)
may not be detectable
by {\it Kepler.} This eliminates our ability to identify and study
 these white dwarfs,
but it also eliminates the possibility that these white dwarfs will
be mistaken for transiting planets. 

The two red circles shown in this panel
correspond to 
KOI-74 (on the left) and KOI-81 (on the right and slightly higher).
Although lensing has not been considered in the light-curve
modeling of KOI-74 and 
KOI-81 carried out by R2010 and vK2010, there was no indication
that it was needed. And clearly 
lensing did not cancel the effect of the transits, since 
{\it Kepler} {\sl did} detect these events. 
The position of KOI-74 on the graph of $M_{WD}$ versus $log(a/{\rm AU})$
indicates that $R_E/R_{WD}$ is expected to be small, so that lensing
should not be significant. In fact, the radius of KOI-74b is larger
than typical of a white dwarf of this mass (R2010), indicating that it may
still be cooling.  We find $R_E/R_{WD}$ is approximately $0.1.$
The position of KOI-81 on the graph of $M_{WD}$ versus $log(a/{\rm AU})$
indicates that, if the radius of KOI-81b satisfies the mass/radius
relationship for white dwarfs, lensing effects may be important.
R2010 find, however, that the radius of KOI-81b is $0.115\, R_\odot,$
placing it near the top of the cooling curve. The value of 
$R_E/R_{WD}$ is $\sim 0.08,$ indicating that the role of
lensing is minimal.    
The systems in the fourth panel have $R_E/R_{WD}< 0.4.$ 
They each have a high probability of generating transits,
but lensing effects will be more subtle and difficult to detect. 
Table 1 summarizes the numerical results for these four categories across
simulations.

\noindent {\bf General Features:} 
We note that, although the majority of white dwarfs orbiting
main-sequence stars lie in 
larger orbits, and are
descended from progenitors that did not interact, 
white dwarfs in wide orbits contribute
little to either transits or lensing events. We expect
only roughly $2$ {\it Kepler} lensing events to be generated by
 white dwarfs
that are the remnants of stars that do not fill their Roche lobes.
The lensing signal will be dominated by white dwarfs formed
in mass transfer binaries. Of these, common envelope survivors  seem likely to 
dominate. This is because, for any given orbital separation,
 white dwarfs emerging from a common envelope tend to have higher mass.  
We note that we also kept 
track of low-luminosity companions to monitored main-sequence.  
We compute that those with orbital periods larger than $\sim 4$~days 
producing transits of {\it Kepler}-monitored stars may be 
$\sim 20$ times less common than the \wdf s which transit \kep\
targets (see Figure 5). These will produce transits with no evidence of
lensing. 

\noindent{\bf Magnification without Transits:} 
The disk of a small object must directly overlap the disk of 
its stellar companion to decrease the  
amount of  light we receive from it. 
Detectable lensing can occur, however,  even without overlap
between the object and its companion. Thus, the cross section for lensing events is higher, and
the lensing events caused by transiting white dwarfs will be supplemented by lensing events   
caused by   
the dwarfs making more distant approaches. The rate of pure lensing events depends on
finite source size effects, 
so it must be computed in each case. The enhancement of a factor will
be  modest. For objects more compact and more massive than white dwarfs, lensing is by far the
dominant effect, and finite-lens effects are not important. (See below.)

\noindent{\bf Neutron Stars and Black Holes:}    
We included more massive stars 
in our simulation, and kept track of those binaries in which
a neutron star or black hole orbits a main sequence star within
roughly $0.5$~AU.
The evolutionary 
pathways that produce such orbits are far less
certain than the evolutionary channels producing close 
white-dwarf/main-sequence binaries. Nevertheless, we know that systems 
with main sequence stars with mass in the range $0.8-2\, M_\odot$
orbit neutron stars and black holes. When the orbits are 
close enough (with semimajor axis of a few solar radii), 
such binaries are detected as low-mass x-ray
binaries. In our simulation   
we considered all binaries with primary 
mass larger than $8.5\, M_\odot$ (the lowest mass star
producing a neutron star remnant) 
and secondary mass between $0.8\, M_\odot$ and $2\, M_\odot.$
We then assumed that $\sim 1/6-1/5$ of the initial orbital separations
could be compatible with evolutionary channels that produce close
neutron-star/main-sequence pairs (see, e.g., Kalogera \& Webbink 1998;
Kiel \& Hurley 2006).   
Although we did not follow the detailed evolution of individual systems,
our approach yielded a binary fraction of neutron stars in close orbits
with main-sequence stars or their remnants of 
($< 0.5\, AU$)  of $5\times 10^{-5}.$ 
We found that {\it Kepler} could detect $1-2$ lensing ``events'' 
by neutron stars or black holes corresponding to approaches close enough
that the compact object transits the companion star. 
In most cases, the
neutron star or black hole would be too dim for 
eclipses of it to be detected.

\section{Conclusion}

\subsection{Results}

We started by considering evolutionary models for KOI-74 and KOI-81.
Our results agree with those of vK2010, which suggest that each system
may have evolved to its present state through a process of stable mass
transfer. Our approach differs from theirs by using a different prescription
for the radius of the donor star prior to envelope exhaustion. Whereas
vK2010 used a formula for the radius which depends on only the core
mass of the donor, our formula also incorporates the dependence on its
equilibrium stellar mass $M_\ast,$  prior to the last phase 
of mass transfer. This has two effects. First our approach tends
to predict
somewhat smaller values of the white dwarf mass. This will be tested
through radial velocity measurements, and Equation (1) can be adjusted
if needed, based on the observations. 
Second, it allows the {\it Kepler's} period measurement  
to constrain both the white dwarf mass and the mass of the donor, thereby
providing useful input for detailed evolutionary models.

We also take seriously the possibility that the masses of KOI-74b and
KOI-81b are in the brown dwarf range, so that the formalism we and vK2010
have applied is not valid. We find an interesting new evolutionary 
pathway. In this case, there is a third body that pumps up the    
eccentricity of an inner binary. Mass transfer from the primary to the 
secondary occurs at periastron, in analogy to the mass transfer process in  
high-mass x-ray binaries. As the orbit evolves, eventually the primary
experiences a dynamical time scale instability at periastron, leading
to a common envelope phase. During the common envelope, the orbit
circularizes and the core of the primary spirals closer to its
main-sequence companion, but avoiding a merger.
The signature
that such an evolution has occurred is a white dwarf in a wider orbit than
expected had there been either stable mass transfer of a common envelope 
initiated from a circular orbit. We do not know if this scenario
is needed to explain either KOI-74 or KOI-81, but if it occurs
in nature, {\it Kepler}
may find binaries that have evolved in this way.       

Whatever the nature of KOI-74 and KOI-81, their discovery inspires us
to compute the frequency of transits by white dwarfs in binaries
that have experienced mass transfer. In many of these cases, the
companion star has gained mass from the progenitor of the white dwarf
and may be considered to be a field blue straggler. 
We have done a set of preliminary  calculations to estimate the
number of such white dwarfs that transit stars in the {\it Kepler}
field. Although our model may be viewed as a ``toy model'' it 
allows us to parameterize the uncertainties involved in the 
complex physics needed to predict the details of the binary evolutions.
We find that, under a wide range of reasonable assumptions, roughly
$0.0025-0.0075$ of systems of the type monitored by {\it Kepler} should
exhibit white dwarf transits. It would be surprising if fewer than $100$
transiting \wdf s were discovered by \kep , while it is possible that
more than $1000$ will be found.  
Furthermore, evidence of gravitational lensing is expected in $10-20\%$
of the \wdf\ transits; anti-transits may be observed in $1/3-1/2$
of the cases in which there are lensing signatures.\footnote{More detailed
calculations are needed to provide reliable estimates of the 
fraction of events exhibiting various lensing signatures. Here we
assume that lensing can be deduced from model fits in events with
$R_E/R_{WD}> 0.8$, and in roughly half of the events
with $0.4 < R_E/R_{WD}< 0.8$, with $\sim 1/2$ of these exhibiting
anti-transits.   
} 
The ubiquity of lensing by the white dwarfs in the \kep\
sample can be understood in terms of mesolensing, the high probability
of lensing associated with certain astrophysical systems, even though
the total mass in lenses, as measured by the lensing optical depth,
is low. (See Di\thinspace Stefano 2008a, 2008b.)   

Our results can be compared with a calculation of the
number of white-dwarf transits expected, conducted before
the \kep\ targets were selected (Farmer \& Agol 2003). 
This previous work had to anticipate the characteristics
of the stars that would be monitored and the
detectability of transits and lensing. Farmer \& Agol (2003) concluded
that at least $50$ transiting \wdf\ binaries could be unambiguously
identified. Furthermore, if detectability issues would be
less problematic than they assumed for their set of target stars,
the number could be increased to $500.$ Because the \kep\ targets
have been specifically selected to make planetary (hence \wdf )
transits detectable for $\sim 90\%$ of the monitored stars,
the higher number they derived seems more likely.
Furthermore, the mass spectrum they assumed for monitored stars was
broader, meaning that the efficiency for producing white-dwarf/main-sequence
binaries is lower.  
Taking this into account, their results can be scaled to produce
rough agreement with the calculations here. We emphasize that
our calculations, as well as those of other population synthesis
simulations, have uncertainties that can best be understood by
exploring the parameter space, as described in \S 3, and summarized
in Table~1.    

\subsection{Directions for Future Work}

It is possible that KOI-74 and KOI-81 are the first examples of
transiting \wdf s. The fact that they are both in close orbits
is expected, since the majority of \wdf s whose transits will be
detected are in close orbits (Table 1). 
They both appear to produce pure transits, without obvious signs of
lensing. (See the discussion in 
\S 4.) Their discovery, and the calculations above suggest several
productive lines of research.  

\begin{itemize}

\item{}  
There should be several, perhaps dozens of anti-transits  
in the data already taken. If these can be identified and
modeled, we will have the first evidence of binary self-lensing and 
the first lensing measurements of the gravitational masses of white dwarfs. 

\item{}
Mass measurements of all transiting objects are crucial.   
Radial velocity measurements should be obtained whenever possible.
Light curve fitting will also be helpful. If enough systems can be
subjected to both types of analyses, we may develop more
confidence in results based on light curve methods alone. When
fitting the light curves, 
it is important to employ models that include the possible
effects of
 lensing.
Although there are potential degeneracies in fits that
include a variety of effects (such as lensing, tides, Doppler boosting),
mathematical 
methods similar to those that test for degeneracy in
 lensing events can be 
applied to transits and anti-transits 
(Di\thinspace Stefano \& Perna 1997).
The result should be 
to
identify those events those that may be associated with compact objects
and to derive the radius and  mass of the compact object. 

\item{} Transiting \wdf s can be subject to additional observations, 
using the opportunities presented by transits to study their 
atmospheres. The radiation from hot white dwarfs entering or leaving
eclipse can also
be used to probe the atmospheres of their companions.   

\item{} 
Lensing by \wdf s will provide unique opportunities to probe the surface
of the stars they orbit.
  
\item{} 
Although we cannot yet predict the orbital or mass distributions of planets, 
binary evolution allows us to predict general 
characteristics of the orbital 
and mass distribution of white dwarfs in close ($a<$ a few AU) binaries.
A set of more detailed calculations is needed,
 to generate individual binaries for each set of
simulations, and to compute the mass transfer history,
the time at which mass transfer ended, 
the radius
of the white dwarf, the light curves, and the characteristics of the 
monitored star (rotation, metalicity) that may have been influenced
by accretion. 

\item{} By discovering orbiting white dwarfs with lower mass than expected
for their measured orbital periods, \kep\ can establish the
contribution of three-body interactions to the formation of blue
stragglers.

\item{} Once a large ensemble of \kep\ events exists, we will be able
to compare their properties with those predicted by the theoretical
work, and learn about a wide variety of outstanding issues,
such as the fraction of matter retained during stable mass transfer
to a main sequence star and the efficiency of common envelope
ejection in a variety of situations.
 
\item{} Many of the mass transfer products in which a white dwarf
orbits a main-sequence star will experience future epochs of
mass transfer onto the white dwarf. Some of these white dwarfs are
destined to become Type Ia supernovae.
Significant uncertainties in our 
understanding of the
progenitors (Di\thinspace Stefano 2010; Di\thinspace Stefano \& Nelson 1996) 
will be addressed by the \kep\ results. 
The same is true for other outcomes, such as
 accretion-induced
collapse. The \kep\ data on the endpoints of the evolution of the
primaries will provide, for the first time, a large number
of reliable starting points for us to compute the
next phase of evolution in close white-dwarf binaries.  
 
\end{itemize} 

\bigskip

\noindent{\bf Implications:}
\kep\ was developed as a 
mission to discover planets. We find that, in addition, it will be a unique
resource to study \wdf s and mass transfer. The results will touch
on stellar evolution, binary evolution, and the formation of intriguing
systems, such as the progenitors of Type Ia supernovae and 
accretion-induced collapse.  

\smallskip
\noindent{\bf Acknowledgements:} It is a pleasure to thank 
Alison J.\ Farmer, Robert J.\, Harris, David Latham,
Hagai Perets, Darin Ragozzine, Jason Rowe, Kailash Sahu, 
Guillermo Torres, and participants in CfA's 
``exoplanet pizza lunch'' for useful discussions. This work was 
supported in part by NSF under AST-0708924 and AST-0908878,
 and by a research and development grant from the CfA.   

\bigskip

\centerline{\bf References} 

\noindent 
Agol, E.\ 2003, \apj, 594, 449 

\noindent 
Batalha, N. et al.\, 2010, arXiv:1001.0349v1
 
\noindent 
De Marco, O., Shara, 
M.~M., Zurek, D., Ouellette, J.~A., Lanz, T., Saffer, R.~A., 
\& Sepinsky, J.~F.\ 2005, \apj, 632, 894 
 
\noindent 
Di\thinspace Stefano, R.\ 2010, ApJ, in press (arXiv:0912.0757)
    
\noindent 
Di\thinspace Stefano, R.\ 2008a, \apj, 
684, 46  
 
\noindent 
Di\thinspace Stefano, R.\ 2008b, \apj, 
684, 59 
 
\noindent 
Di\thinspace Stefano, R., \& Perna, R.\ 1997, \apj, 488, 55 
 
\noindent 
Di\thinspace Stefano, R., \& Nelson, L.~A.\ 1996, Supersoft X-Ray Sources, 472, 3 

\noindent 
Eggleton, P.~P.\ 2002, \apj, 
575, 1037 

\noindent 
Eggleton, P.~P.\ 1986, private communication  

\noindent 
Eggleton, P.~P.\ 1983, \apj, 
268, 368 

\noindent 
Eggleton, P.~P., \& Kisseleva-Eggleton, L.\ 2001, \apj, 562, 1012 

\noindent 
Fabrycky, D., \& Tremaine, S.\ 2007, \apj, 669, 1298 

\noindent 
Farmer, A.J. \& Agol, E. 2003, 592, 1151

\noindent 
Harris, H.~C., et al.\ 
2008, \apj, 679, 697 

\noindent 
Kalogera, V., \& Webbink, R.~F.\ 1998, \apj, 493, 351 

\noindent 
 Kiel, P.~D., \& Hurley, J.~R.\ 2006, \mnras, 369, 1152 

\noindent 
 Kozai, Y.\ 1962, \aj, 67, 591 

\noindent 
Landsman, W., 
Aparicio, J., Bergeron, P., Di Stefano, R., 
\& Stecher, T.~P.\ 1997, \apjl, 481, L93 

\noindent 
Latham, D.W.\, 2010, private communication   

\noindent Lidov, M.L. 1961, Iskusstvennye Sputniki Zemli, No. 8, p. 5 

\noindent 
Mathieu, R.~D., \& Geller, A.~M.\ 2009, \nat, 462, 1032 

\noindent 
 Nelemans, G., \& Tout, C.~A.\ 2005, \mnras, 356, 753 
 
\noindent 
Parsons, S.~G., Marsh, 
T.~R., Copperwheat, C.~M., Dhillon, V.~S., Littlefair, S.~P., G{\"a}nsicke, 
B.~T., \& Hickman, R.\ 2010, \mnras, 61  
\noindent 
Perets, H.~B., \& Fabrycky, D.~C.\ 2009, \apj, 697, 1048

\noindent 
Pyzras, S. et al.\, 2009, MNRAS, 394, 978 

\noindent 
Rappaport, S., 
Podsiadlowski, P., Joss, P.~C., Di Stefano, R., 
\& Han, Z.\ 1995, \mnras, 273, 731 

\noindent 
Rowe, J.F. et al.\, 2010,  arXiv:1001.3420

\noindent 
Sahu, K.~C., \& Gilliland, R.~L.\ 2003, \apj, 584, 1042

\noindent 
Sarna, M.~J., Ergma, E., 
\& Ger{\v s}kevit{\v s}-Antipova, J.\ 2000, \mnras, 316, 84 

\noindent 
van Kerkwijk, M.H. et al.\, 2010,  arXiv:1001.4539

\noindent 
Vidrih, S., et al.\ 
2007, \mnras, 382, 515 

\noindent Webbink, R.~F.\ 2008, 
Astrophysics and Space Science Library, 352, 233

\noindent 
Winget, D.~E., Hansen, 
C.~J., Liebert, J., van Horn, H.~M., Fontaine, G., Nather, R.~E., Kepler, 
S.~O., \& Lamb, D.~Q.\ 1987, \apjl, 315, L77 

\end{document}